\documentclass[12pt]{spieman}  
\usepackage{amsmath,amsfonts,amssymb}
\usepackage{graphicx}
\usepackage{setspace}
\usepackage{tocloft}

\title{CCAT: FYST Prime-Cam Readout Software: A framework for massively scalable KID arrays}

\author[a,*]{James R. Burgoyne}
\author[a]{Adrian K. Sinclair}
\author[b,f]{Scott C. Chapman}
\author[g]{Steve K. Choi}
\author[d]{Cody J. Duell}
\author[c]{Anthony I. Huber}
\author[d]{Zachary B. Huber}
\author[d]{Ben Keller}
\author[d]{Lawrence Lin}
\author[d,i]{Michael D. Niemack}
\author[a]{Douglas Scott}
\author[d,e]{Eve M. Vavagiakis}
\author[d,h]{Samantha Walker}
\author[a]{Matt Xie}
\author[ ]{the CCAT collaboration}

\affil[a]{Department of Physics and Astronomy, University of British Columbia, Vancouver, BC~V6T~1Z1, Canada}
\affil[b]{Department of Physics and Atmospheric Science, Dalhousie University, Halifax, NS B3H 3J5, Canada}
\affil[c]{Department of Physics and Astronomy, University of Victoria, Victoria, BC~V8W~2Y2, ~Canada}
\affil[d]{Department of Physics, Cornell University, Ithaca, NY~14853, USA}
\affil[e]{Department of Physics, Duke University, Durham, NC 27708, USA}
\affil[f]{Herzberg Astronomy and Astrophysics Research Centre, Victoria, BC V9E 2E7, Canada}
\affil[g]{Department of Physics and Astronomy, University of California, Riverside, CA 92521, USA}
\affil[h]{Cornell Center for Materials Research, Cornell University, Ithaca, NY 14853, USA}
\affil[i]{Department of Astronomy, Cornell University, Ithaca, NY 14853, USA}
\affil[j]{Quantum Sensors Division, National Institute of Standards and Technology, Boulder, Colorado~80305, USA}
\affil[k]{Cornell Center for Astrophysics and Planetary Sciences, Cornell University, Ithaca, New York~14853, USA}
\affil[l]{Department of Applied and Engineering Physics, Cornell University, Ithaca, New York 14853, USA}
\affil[m]{Department of Physics, University of Colorado, Boulder, Colorado 80309, USA}

\cftpagenumbersoff{figure}
\cftpagenumbersoff{table} 
\begin{document} 
\maketitle

\begin{abstract}
We outline the development of the readout software for the Prime-Cam and Mod-Cam instruments on the CCAT Fred Young Submillimeter Telescope (FYST), \texttt{primecam\_readout}. The instruments feature lumped-element kinetic inductance detector (LEKID) arrays driven by Xilinx ZCU111 RFSoC boards. In the current configuration, each board can drive up to 4000 KIDs, and Prime-Cam is implementing approximately 25 boards. The software runs on a centralized control computer connected to the boards via dedicated ethernet, and facilitates such tasks as frequency-multiplexed tone comb driving, comb calibration and optimization, and detector timestream establishment. The control computer utilizes dynamically generated control channels for each board, allowing for simultaneous parallel control over all, while uniquely tracking diagnostics for each. 
This work demonstrates a scalable RFSoC readout architecture where computational demands increase linearly with the number of detectors, enabling control of tens-of-thousands of KIDs with modest hardware, and opening the door to the next generation of KID arrays housing millions of detectors.
\end{abstract}

\keywords{detector readout, kinetic inductance detector, control systems, astronomy software, astronomical instrumentation, millimeter astronomy, astronomical detectors}

{\noindent \footnotesize\textbf{*}James Burgoyne,  \linkable{jburgoyne@phas.ubc.ca} }

\begin{spacing}{1}   

\section{Introduction}
\label{sec:introduction}

Submillimeter astronomy ($0.3\mbox{--}1$ mm) constitutes an important domain within observational astrophysics, with applications in topics such as the Epoch of Reionization (EoR), galaxy evolution on cosmic timescales, measuring CMB foregrounds, Galactic magnetic field polarization and morphology, and large-scale structure evolution through the Sunyaev Zeldovich (SZ) effect and recombination epoch Rayleigh scattering \cite{CCAT-PrimeCollaboration2023}.
Low latency submillimeter detectors would also unlock research into time-domain phenomena such as very energetic transients, including supernovae, $\gamma$-ray bursts, X-ray binaries, merging neutron stars, and tidal disruption events \cite{CCAT-PrimeCollaboration2023}.

Yet, the pursuit of astronomical knowledge within this spectral domain entails a series of formidable challenges, chiefly originating from technological and environmental constraints. 
These challenges include the necessity for highly sensitive detectors, precise thermal regulation of instruments, and the amelioration of atmospheric interference, all of which necessitate state-of-the-art methodologies and techniques. 

For sub-millimeter telescope cameras, achieving high sensitivity and fast mapping speeds has traditionally relied on transition-edge sensor (TES) bolometers. 
These detectors operate at cryogenic temperatures, offering excellent sensitivity through their sharp transition from superconducting to normal states upon absorbing a photon. 
However, TES detectors face limitations. 
Their readout requires complex electronics with intrinsic noise sources such as amplifier noise and crosstalk. 
Additionally, the need to individually read out each pixel restricts large-scale arrays, hindering mapping efficiency. While TES bolometers remain a powerful tool for sub-millimeter astronomy, the push for even greater detector counts and reduced fabrication complexity is driving the exploration of alternative detector technologies \cite{Niemack2016}.

Kinetic inductance detectors (KIDs; alternately microwave kinetic inductance detectors, MKIDs) have shown great promise for the development of high-performance detector arrays in submillimeter and millimeter astronomical instrumentation applications \cite{Day2003}. 
A KID consists of a superconducting LC circuit, which forms a microwave resonator whose resonant frequency changes with the absorption of incident photons. 
KIDs act as high quality-factor notch filters and can therefore be frequency-multiplexed, allowing thousands of detectors to be operated on the same radio-frequency (RF) network, and enabling large detector arrays \cite{Baselmans2017}. 
As superconducting resonators, KIDs have already demonstrated photon noise-limited sensitivity performance, which is essential for many demanding astronomical applications \cite{Gusten2006}.
As a result of their high multiplexing factors, KIDs offer significant gains in mapping speed while also being simpler to fabricate than comparably- sensitive TES detectors.

However, much of the simplification of hardware comes at the cost of more complex readout and analysis software and firmware.
KID readout techniques rely on measuring the phase and amplitude changes of an RF tone placed on resonance.
A tone `comb' is generated consisting of, potentially, thousands of tones, with each tone frequency tuned to a specific KID in the array.
The phases and amplitudes of these tones are then altered by their respective KID in such a way that the photon flux can be determined.
The readout for such a system needs to be capable of generating and measuring phase-coherent signals fast enough and on the scale of the size of the detector arrays required for the particular application \cite{Esteras2014}.

\section{Telescope and Camera Overview}
\label{sec:overview}

\begin{figure}
\includegraphics[width=\textwidth]{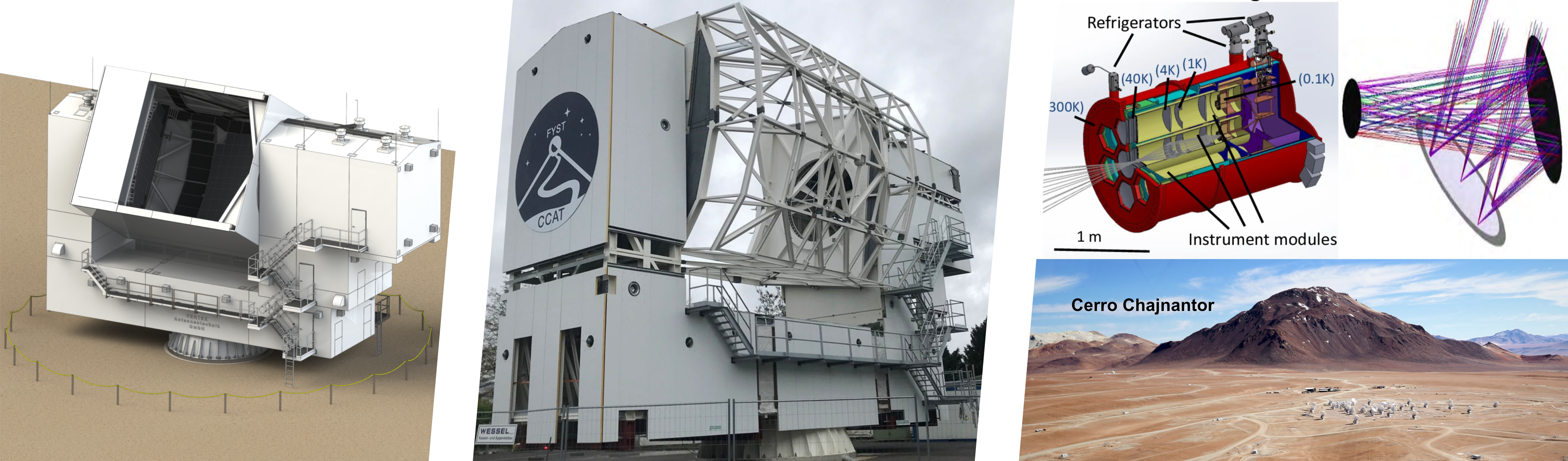}   
\caption{
CCAT/FYST computer render (left), and during assembly (center). Prime-Cam and telescope optical path (top-right). Observatory site at the top of Cerro Chajnantor (bottom-right), with ALMA in the plateau below.}
\label{fig:telescope}
\end{figure} 

The Prime-Cam instrument \cite{Vavagiakis2018, Choi2019}, on the CCAT Collaboration's Fred Young Sub-millimeter Telescope (FYST, pronounced “feast”) \cite{Parshley2018, Parshley2018a}, is a multi-module cryostat that will function as a primary instrument when it is deployed in 2026. 
FYST will be a high-efficiency, wide-field telescope, featuring a 6-m primary with a surface accuracy of 10 microns and an 8$^\circ$ field-of-view (FoV) at 3 mm, in a crossed-Dragone design. 
The observatory is situated at 5600 m elevation on Cerro Chajnantor in the Atacama desert of northern Chile. 
FYST and Prime-Cam have been developed for studies that involve large-scale, high-sensitivity polarimetric, photometric, and spectroscopic mapping \cite{CCAT-PrimeCollaboration2023}. 
They have been designed for very deep mapping through the telluric windows from 100 to 900 GHz (3 to 0.33 mm wavelength). 
Surveys undertaken with FYST are designed to surpass previous surveys (e.g. with CSO, JCMT, APEX, SPT and ACT, Planck, Herschel) in terms of confusion-limited depth, and areal and/or frequency coverage \cite{CCAT-PrimeCollaboration2023}. 

Prime-Cam is a single large sub-kelvin cryostat capable of supporting seven instrument module spaces. Each instrument module intercepts a part of the field-of-view, and is independent of the others such that they can be swapped out. As such, each module contains its own optics, filters, and detectors \cite{Vavagiakis2022}. 
All modules will be filled with lumped element kinetic inductance detectors (LEKIDs), chosen to enable scaling to the pixel-pitch and detector counts required to accomplish CCAT science goals. \cite{CCAT-PrimeCollaboration2023}.
All module arrays are fabricated by the Quantum Sensors Division at the National Institute for Standards and Technology (NIST) in a state-of-the-art manufacturing process involving multiple etching passes to achieve 95\%+ resonator yield, with focus given to both high quality-factor and low resonant frequency collision count.
The LEKIDs are driven by Xilinx ZCU111 boards containing the Zynq UltraScale+ Radio Frequency System-on-Chip (RFSoC). 

Since the modules contain similar detectors and readouts, a common readout software, \texttt{\-primecam\_readout}, and readout firmware, \texttt{\-primecam\_gateware}, were developed. 
Details of the latter can be found in Sinclair et al 2024.
This paper describes the development and state of the former.

\section{Software Requirements}

The readout firmware \cite{Sinclair2022}, running on the ZCU111 reconfigurable logic fabric, and the readout software, running on the ARM processor, work in tandem to ultimately generate the calibration data and timestreams for each resonator. The software is designed to meet five key requirements. 
\begin{enumerate}
\item \textbf{Firmware attachment:} Connect to the firmware on each board to provide control of tone comb generation and other firmware-level functionality.
\item \textbf{Tone comb management:} Provide the higher-level functionality necessary to produce and calibrate the tone combs on each board, including analysis algorithms (e.g., identifying resonators in a full bandwidth frequency sweep).
\item \textbf{Multi-board coordination:} Coordinate and manage the operation of all the boards simultaneously.
\item \textbf{Prime-Cam integration:} Seamlessly integrate with the Prime-Cam Control Software (PCS).
\item \textbf{Customization:} Allow for the addition of user-defined functionality, especially for testing and diagnostics.
\end{enumerate}

\section{Architecture}
\label{sec:architecture}

\begin{figure}
\includegraphics[width=\textwidth]{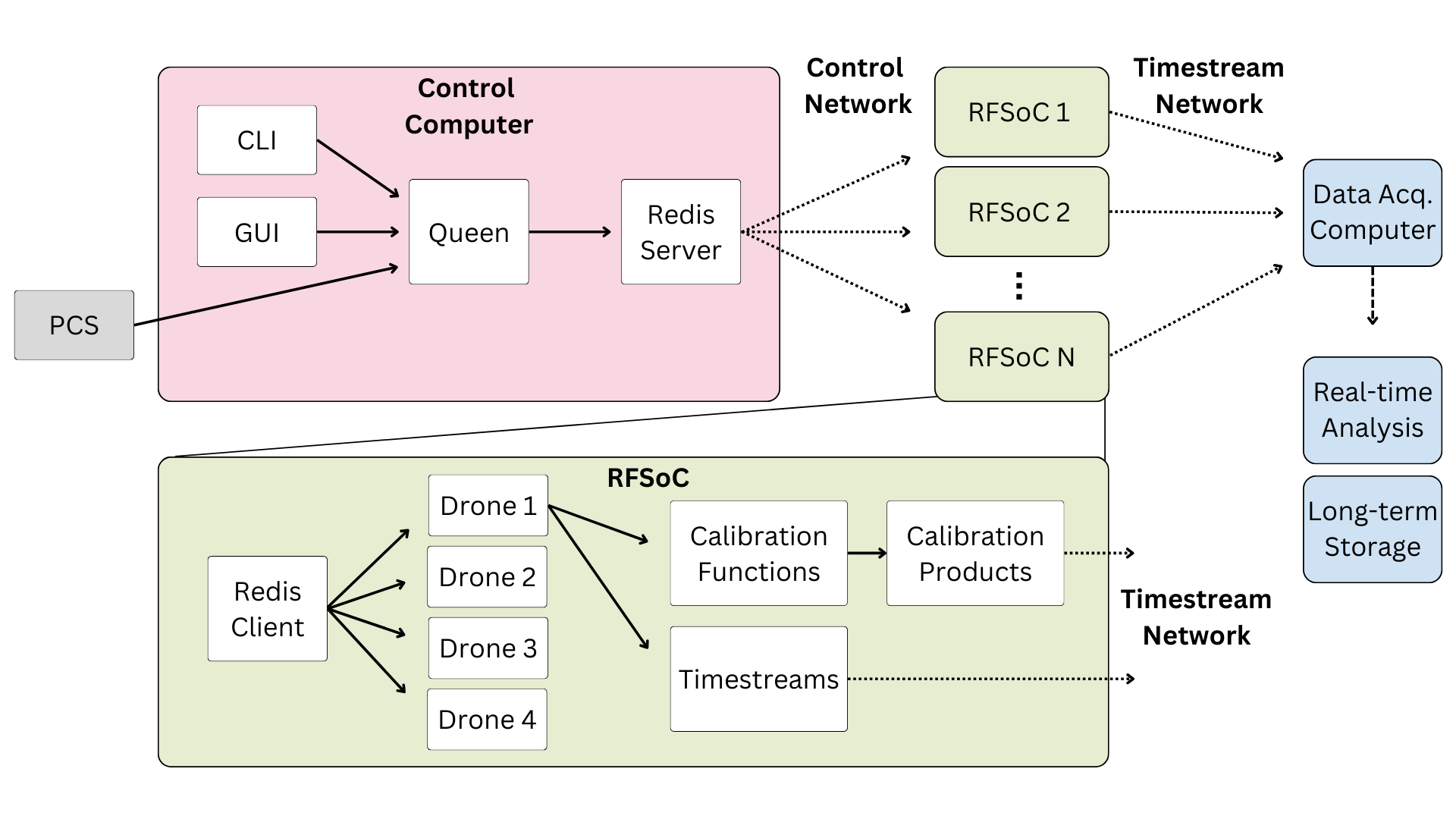}   
\caption{
Overview of the readout software, hardware, and network architectures. The pink box represents software running on the control computer hardware. The large green box represents the software running on one of the boards. The small green boxes indicate that there are many parallel boards. The blue boxes represent the data acquisition and analysis hardware that collects and processes the timestreams from the boards. The gray box represents the Prime-Cam Control Software, which interfaces with the observatory control software.} 
\label{fig:hardware_architecture}
\end{figure} 

An overview of the readout hardware and software architecture can be seen in Fig.~\ref{fig:hardware_architecture}.
The primary components are the control computer, the control network, the parallel and independent RFSoC boards, the data acquisition (timestream) network, and the data acquisition (storage) computer.
The control computer mediates tasks from the Prime-Cam Control Software (PCS), passing commands to the boards via the control network, and receiving relevant status and diagnostic data back from them.
PCS is evolved directly from Simons Observatory observatory control software \cite{}.
Each board is responsible for a segment of the LEKID array, and is able to drive up to 4$\times$1000 KIDs with our current algorithms and the bandwidth constraints imposed by the hardware.
Each resonator produces a timestream with two components: the in-phase ($I$) and quadrature ($Q$).
This decomposition represents the signal's projections onto orthogonal cosine and sine carriers with the carrier's center frequency. 
In polarization-sensitive instrument modules (e.g. the 280-GHz and 850-GHz modules) each physical pixel contains two resonators sensitive to orthogonal polarizations, and there are numerous orientations across the arrays and modules.

Each RFSoC board is equipped with an ARM Cortex processor and is configured to run the PYNQ image v2.6, which incorporates a Linux filesystem and a Python API. 
PYNQ is an open-source framework from AMD that facilitates embedded system design using Xilinx Zynq devices by leveraging Python, Jupyter notebooks, and Python libraries to target the programmable logic and microprocessors within Zynq platforms. 
PYNQ enables the creation of high-performance embedded applications for real-time signal processing, video processing, and hardware-accelerated algorithms.
This integrated setup facilitates seamless interaction with the FPGA gateware and other hardware peripherals. 
We also leverage the built-in microprocessors to function as independent computing nodes responsible for the bulk of software computations. 
Overall this arrangement resembles a distributed computing network and offers a high degree of scalability while eliminating the need for a high-performance user terminal to handle multiple simultaneous computation tasks. 
Additionally, it reduces the data transmission requirements between the user terminal and each RFSoC.

During regular operation, only key-value pairs are necessary for transmitting command strings and status bits. 
\texttt{Redis}, an in-memory database and messaging service, employs TCP (Transmission Control Protocol) to ensure reliable message exchange between clients and the server in a client-server model. 
Communication between clients and \texttt{Redis} utilizes the RESP (REdis Serialization Protocol), a text-based protocol specifically designed for efficient data exchange between \texttt{Redis} and applications.
Despite being primarily written in C for high performance, \texttt{Redis} features Python bindings that we have seamlessly integrated with the PYNQ framework. 
Its inherent messaging queue capabilities, allowing publishing and subscribing to various channels, make it an ideal choice for simultaneously commanding the RFSoCs over the network. 

Each detector's time-ordered data (TOD) are streamed from the RFSoC boards using User Datagram Protocol (UDP) on a dedicated,
high-bandwidth network infrastructure (the `timestreams'). 
UDP is a connectionless protocol; it prioritizes speed over reliability by forgoing handshakes and acknowledgments. 
This makes it ideal for real-time data transfer where minimal latency is crucial, such as in high-speed detector applications. 
In essence, the RFSoC continuously transmits a high-volume data stream (a `fire hose' of $\sim 128$ Gb/s) to pre-configured IP addresses, regardless of whether a receiver is present. 
This approach prioritizes data delivery speed at the expense of error checking, making it suitable for scenarios where real-time data are essential and missing a few packets is less critical than incurring delays. 

Each detector generates a timestream consisting of large 8250-byte jumbo frames for efficient data transfer. 
Each frame incorporates a well-defined header containing essential metadata such as endianness, data description, and the destination's MAC address for proper routing. 
Notably, packet generation is entirely offloaded to the RFSoC's field-programmable gate array (FPGA) fabric, eliminating processor intervention. 
This approach leverages the inherent parallelism and hardware acceleration capabilities of the FPGA. 
Compared to software-based packet generation on a processor, hardware coding within the FPGA fabric offers significant performance advantages. 
FPGAs execute instructions concurrently, enabling high-throughput data processing and packet formation at line rates. 
Additionally, hardware avoids the overhead associated with context switching, memory access, and instruction fetching that plagues processors. 
This offloading significantly reduces latency and improves overall system efficiency, allowing the processor to focus on higher-level tasks like data analysis and control.

The resultant timestream packets are captured by a module-dedicated data-acquisition computer. 
The data-acquisition computer is primarily responsible for converting the raw packets into the \texttt{g3} format and then forwarding to quick-view computers and on-sight storage computers that subsequently upload it to overseas data centers.
The \texttt{g3} format, developed by the South Pole Telescope, is being extended by Simons Observatory and CCAT for use with our system to take advantage of the structure of the underlying \texttt{spt3g} software file format \cite{Koopman2020}.

\section{Signal Pipeline}

This section details the signal generation and processing pipeline within the readout system, focusing on maximizing signal-to-noise ratio (SNR) while accommodating a large number of tones.

\subsection{Digital Waveform Generation}

\begin{figure}
    \includegraphics[width=\textwidth]{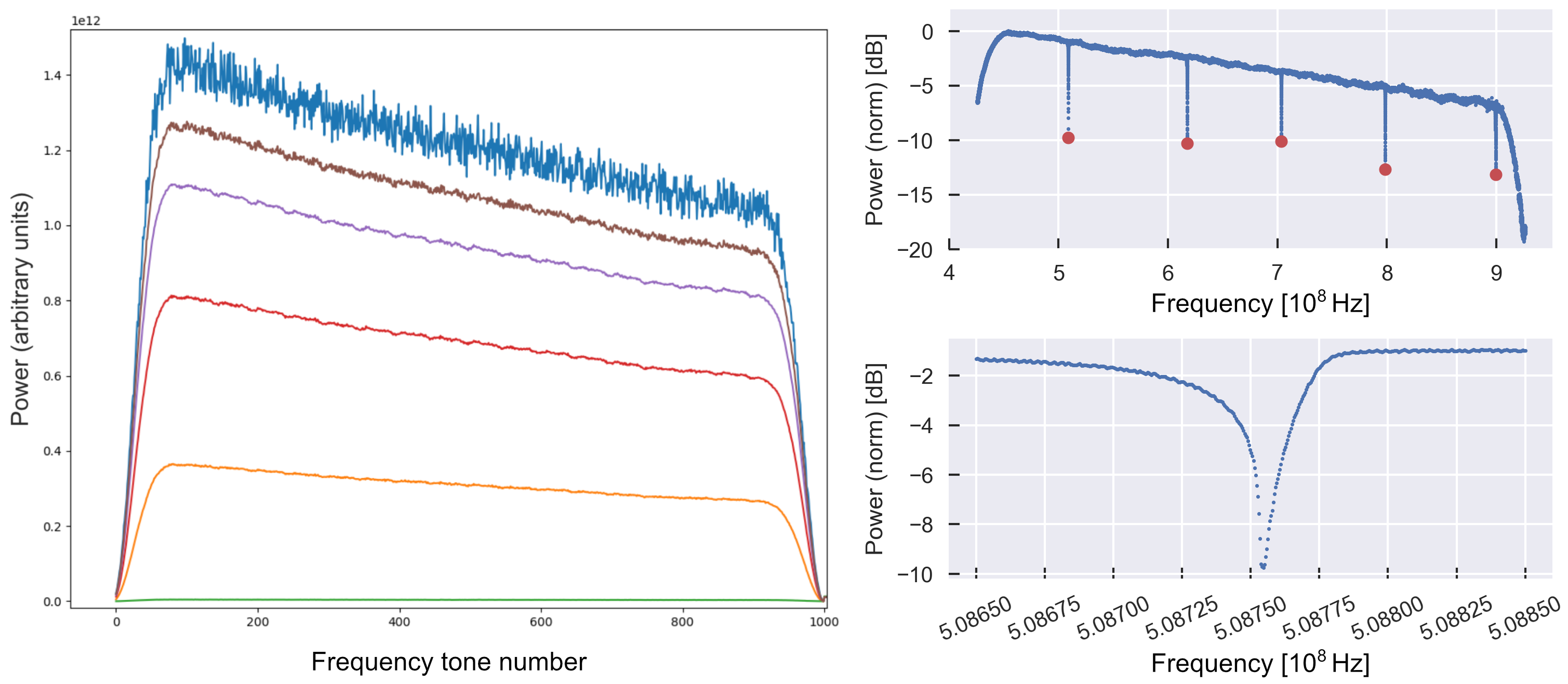}   
    \caption{\\
    \textbf{(Left)} The generated tone comb as produced by this system, in loop-back mode, where each tone power is a data point with a frequency tone number. The main feature is the characteristic shape of the output from the DAC. The same tone comb is being shown here with varying tone powers. Note that all the tones are generated with the same power (per color), and variations are due to the generation process. The extreme variations as power increases are due to going beyond the power that the DAC is capable of producing.\\ 
    \textbf{(Top-right)} A VNA sweep (full bandwidth sweep) produced by this system showing five resonators in an early test array. The red dots indicate the positions that the system has identified resonators.\\
    \textbf{(Bottom-right)} A target sweep produced by the system of a single resonator from the VNA sweep.} 
    \label{fig:sweeps}
\end{figure}

The software constructs a digital representation of the desired `tone comb' waveform, which comprises the sum of a user-defined set of tones, initially distributed uniformly across the available bandwidth (currently 512 MHz). 
Subsequently, these tones are precisely targeted to specific resonator frequencies for optimal coupling (see Fig \ref{fig:sweeps} for an example of a tone comb). 
The software algorithm responsible for generating the tone comb waveform adheres to several crucial constraints. 
The primary objective is to produce the user-defined set of target tone frequencies. 
To maximize the SNR, the waveform must achieve the highest possible amplitude while adhering to the peak voltage limitations of the DAC. 
This necessitates meticulous manipulation of individual tone amplitudes and phases, a task handled by a dedicated software algorithm focused on crest factor optimization.
Following software generation, the digital tone comb waveform is transferred to a random-access memory (RAM) unit. Subsequently, the firmware, operating in conjunction with a user-defined local oscillator (LO) setting, utilizes this waveform data to generate an analog replica via the on-board DAC. This analog signal is then injected into the instrument's Radio Frequency (RF) network. This process is repeated for each of the instrument's RF networks (up to four in the current iteration).

\subsection{Signal Processing and Digitization}

To optimize noise performance, the signal is attenuated by a software-controlled variable attenuator (integrated within the tone power optimization system) before entering the cryostat. 
The generated tone comb traverses the instrument and interacts with the detector arrays, leaving signal modifications from both the signal path and electronics, as well as the intended attenuation from the resonators.
Following attenuation, a low-noise amplifier (LNA) amplifies the signal prior to its return for analog-to-digital conversion (ADC). 
The digitized signal is then demodulated into in-phase ($I$) and quadrature ($Q$) components, resulting in two independent time series. 
These time series are subsequently transmitted to a dedicated data acquisition computer, which performs two primary functions: conversion of the raw $I/Q$ data streams into G3 formatted files, a standardized format for storing such data; and dissemination of the G3 files to quick view map making computers for time-domain science and long-term archival storage.

\section{Networking}

The readout system utilizes two dedicated ethernet networks for distinct communication purposes.

\begin{enumerate}
\item \textbf{Control Network:} This network facilitates communication between the Control Computer and the RFSoC boards using the \texttt{Redis} Serialization Protocol (RESP) over TCP. 
It employs a publish-subscribe (pub/sub) messaging pattern using \texttt{Redis}, a high-performance in-memory data store. The Control Computer acts as a \texttt{Redis} server, publishing commands as strings on dedicated channels. 
Each command has a unique channel, and if applicable, can be targeted towards specific subsets of drones (sub-sections of the \texttt{primecam\_readout} software controlling individual RF networks). 
Simultaneously, corresponding return data channels are generated. 
The RFSoC boards subscribe to relevant channels, receiving commands and publishing their responses. 
This decoupled pub/sub architecture allows for efficient and scalable communication, while providing an architecture that is highly robust to failures.
\texttt{Redis} offers extreme efficiency and low computational burden on the Control Computer, enabling massive scalability – potentially millions of KIDs – without significant control system resource limitations or added system complexity.
\item \textbf{Timestream Network:} This network provides for high-throughput data transfer from the RFSoC boards to the Data Acquisition (DAQ) computer. The boards continuously transmit a `fire hose' of user datagram protocol (UDP) packets containing the raw KID $I$ and $Q$ data. The Timestream Network is designed for high bandwidth to accommodate this continuous data stream.
\end{enumerate}

The key advantage of this dual-network approach lies in its separation of concerns. The Control Network efficiently manages communication and commands, while the Timestream Network focuses on raw data transfer. This segregation avoids network congestion and ensures smooth operation for both control and data acquisition functions.

Scalability is further enhanced by the distributed nature of the DAQ system. Each DAQ computer manages a specific number of RFSoC boards, limiting its workload. As the number of KIDs increases, additional DAQ computers and networks can be seamlessly integrated, expanding the system's capacity, without increasing complexity to any individual part of the system.

Overall, the networking architecture prioritizes efficiency, scalability, and clear separation of communication channels. This design enables the KID camera system to handle massive detector arrays while maintaining reliable control and data-acquisition capabilities.

\section{Driven Waveform Generation}

KIDs employ an actively-driven, non-linear readout scheme. 
This process involves generating and transmitting a tailored waveform optimized for three key factors: 
1) the specific detector array configuration, 
2) the real-time electrical characteristics of the array and transmission path, and 
3) the environmental conditions, particularly atmospheric effects influenced by weather and telescope pointing elevation. 
Solving for the optimal waveform typically involves an iterative feedback algorithm. 
The resulting optimized waveform represents the constructive interference of a set of frequency-multiplexed tones, each targeting a specific resonator within the array. 
Each tone is characterized by three independent parameters, namely frequency, amplitude, and phase.

\subsection{VNA Sweep}

The initial step in determining the optimal readout waveform involves characterizing the instantaneous resonant frequencies of each detector within the array, achieved by first generating a comb waveform encompassing the full operational bandwidth of the resonators (currently limited to a 512-MHz band by the DAC). 
The comb waveform comprises 1000 equally spaced frequency tones of equal amplitude, and an iterative process is used to solve for optimum phases such that the maximum voltage of the DAC is not exceeded.
The center frequency of the comb is established by the local oscillator (LO) frequency. 
Subsequently, the LO frequency is swept across the entire bandwidth with a resolution sufficient to identify every resonator's position. 
This process, known as a `VNA sweep', requires a step size smaller than the expected resonance linewidth but not excessively small, as a finer step resolution increases the sweep time. 
Finally, the modified transmission signal acquired during the VNA sweep is processed using a custom cleaning and peak-finding algorithm to extract the precise frequency locations of all resonators.

\subsection{Target Sweep}

Following the initial VNA sweep that provides a coarse estimate of the resonator frequencies, a refined comb waveform is generated. 
The new comb comprises tones centered on the approximate frequencies identified previously. 
A process analogous to the VNA sweep is then implemented, but with a significantly smaller step size precisely targeted around each resonator's estimated frequency. 
High-resolution sweeps aim to isolate the current resonant frequencies with the maximum useful precision, typically limited to approximately 20 Hz before frequency noise dominates. 
The resonant frequencies are identified as the points within each tone sweep where the signal magnitude exhibits the most significant attenuation. 
Finally, the precisely measured resonator properties are employed to construct the final readout waveform. 
The final comb waveform strategically positions each tone at the most sensitive region of the resonator response on the leading edge. 
Positioning in this way yields the largest frequency shift with power absorbed, leading to maximized sensitivity.  
Critically, it also allows for the greatest possible shift in the resonator's frequency before sensitivity declines due to straying too far from the optimal resonance point.

\section{Calibration and Data Preprocessing}

Calibrating a KID for astronomical observations necessitates establishing a well-defined relationship between the incoming light and the detector's response.
This involves characterizing the responsivity of each pixel through its resonance frequency shift with varying optical load.
Our approach is to place a tone on resonance and monitor the resultant transmission attenuation fluctuations of that tone.
This is then later converted to a signal proportional to incident photon power.
That conversion necessitates understanding the characteristics of the detector during observations.
Those characteristics are derived from the target sweeps outlined above.
If those characteristics change, a new target sweep is performed, and understanding the conditions that lead to this happening are important. 
There are several considerations for this that are outlined below.

\subsection{Non-Linear Response}

The detector's response exhibits a non-linear dependence on  optical power across its operational range. Consequently, linear approximations only provide acceptable uncertainty within a limited regime, with the size of this regime being highly specific to both the detector and its operating environment. Non-linear conversion functions offer a means to expand this regime. We present a novel implementation of a non-linear conversion method that maintains computational complexity comparable to linear approximations (the `$IQ$-angle' method) \cite{Gao2008}. This approach enables extended observational periods before recalibration becomes necessary.

\subsubsection{IQ-angle Method}

The $IQ$-angle method is a novel implementation of the theoretical method (outlined in Gao 2008 \cite{Gao2008}) for calculating the frequency shift, which is proportional to incident power, from the observation $I$ and $Q$ data as well as a calibration target sweep.

The target sweep data, used for characterizing KIDs, typically exhibits a unique minimum in magnitude within a single resonator's response. 
This minimum corresponds to the resonant frequency of the resonator.
The target sweep path in the $IQ$-space ideally resembles a circle and is termed the `$IQ$-loop'. 
Due to the limited frequency range of the target sweep, within a few line widths around the resonant frequency, the $IQ$-loop remains relatively confined with minimal cyclical overlap or deviation from the expected circular path. 
Consequently, a robust estimate of the $IQ$-loop's `center' can be obtained by averaging the minimum and maximum values of both the $I$ and $Q$ components.

The $IQ$-angle method exploits the target sweep data and observed detector response data by first shifting both datasets such that the center of the target sweep's IQ-loop aligns with the origin in IQ-space. This facilitates the calculation of the angle for each data point relative to the origin. Since the target sweep design assigns a specific frequency to each data point, a relationship between the $IQ$-space angle and its corresponding frequency exists. Leveraging an extremely efficient interpolation method, the target sweep data are interpolated at each observed angle to map to the associated frequency.

This method boasts computational efficiency comparable to linear approximations, with conversion accuracy of non-linear methods, and enables extended observation periods before requiring recalibration.
Signal and computation comparisons can be seen in Fig \ref{fig:commonmode}.

\subsection{Temperature Sensitivity}
\begin{figure}
\includegraphics[width=\linewidth]{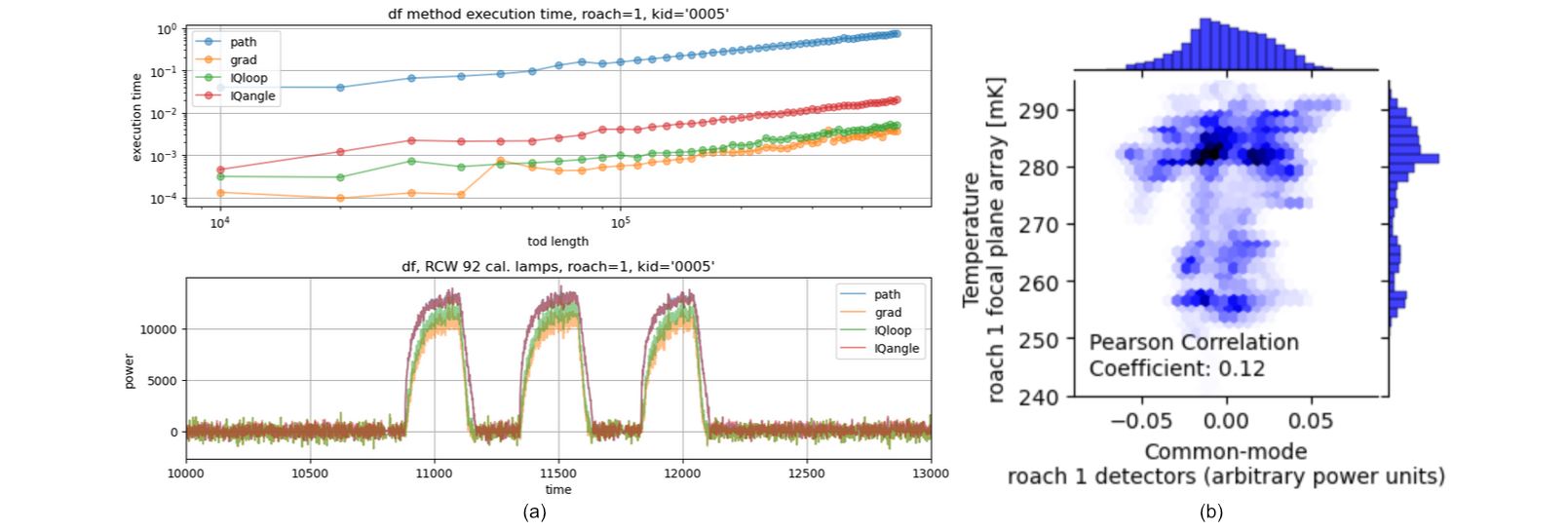}   
\caption{Testing with data from a 20 minute observation of source RCW 92 done with the BLAST-TNG experiment \cite{Galitzki2014}. \\
\textbf{(a)} Frequency-shift algorithm testing. The top panel shows computational time of each of the tested algorithms, and the bottom panel shows a sample of the algorithm output over a small sample (calibration lamp cycling). The algorithm developed in this paper is nicknamed `IQangle', while `IQloop' and `grad' are linear approximation methods, and `path' is a full path integral calculation. Note that the IQangle method retains the full response of the path method, while substantially reducing the computational time required by almost 2 orders of magnitude.\\
\textbf{(b)} Testing done with data from the BLAST-TNG experiment \cite{Galitzki2014} showed a negligible correlation between cryostat temperature fluctuations and the common-mode of the detector signals, demonstrating that KID responsivity precision is dominated by sources other than operating temperature fluctuations. Note that operational temperatures have been heuristically shown to be optimal well above cryostat minimum achievable temperatures, primarily due to reduced TLS noise \cite{Swenson2013}. In this figure the power units are normalized to the calibration lamp peak signal.} 
\label{fig:commonmode}
\end{figure}

KIDs depend on superconductivity, and therefore a cryogenic environment, and have optimal performance inside a small temperature window of order 0.1 K.
However by design they are much less sensitive to bath temperature fluctuations than to optical signals within that window.
The upper end of this window is defined theoretically by the critical temperature of the superconducting material, but practically by thermal energy over-saturating the detector and reducing the quality-factor and sensitivity.
The lower end of this window is due to two-level system (TLS) noise. 
Fluctuations between quantized energy states in microscopic defects at the superconductor-substrate interface, known as two-level systems, generate excess electrical noise that impedes the weak signal detection capabilities of KIDs \cite{Swenson2013, Noroozian2009}.
Despite a decrease in total excited TLS at lower temperatures, extended trapping times in the higher energy state lead to a net increase in TLS noise.
The overall effect is that TLS noise is inversely related to temperature.
These effects mean there is an optimal temperature for each array to operate at, but within a window around that temperature the responsivity is insensitive to temperature fluctuations, as can be seen in Fig \ref{fig:commonmode}(b).

\subsection{Electronics Optimization}
The readout analog transmission is attenuated down as is appropriate to not over-drive the detectors.
Each detector has a unique response, and each tone has a likewise unique amplitude generated in the waveform before the common attenuation.
The unique tone amplitudes are software determined in an iterative process to optimize the SNR.
Since the signal is strongly attenuated before the cryostat the detectors produce very weak electrical signals. 
The readout electronics need to amplify these signals for processing using a low-noise amplifier (LNA). The LNA is situated inside the cryostat to avoid introducing thermal noise.
However, increasing the amplifier gain can also amplify inherent electronic noise within the system. 
Calibration involves finding the optimal gain setting that maximizes the SNR, and is done in a iterative closed-loop process before observations begin.

\subsection{Non-Uniform Responsivity}
Detector responsivity is not uniform across the array, and
each detector must have their sensitivity characterized for the map-making process.
For Prime-Cam, this characterization is empirically derived using laboratory and then ultimately on-sky measurements, rather than calibration lamp measurements in the optical path.

\section{Conclusions}

Kinetic inductance detectors offer significant advantages for submillimeter astronomy due to their high sensitivity and scalability. 
However, unlocking their full potential necessitates complex readout software and data analysis techniques. 
The Prime-Cam instrument on the Fred Young Submillimeter Telescope exemplifies a cutting-edge solution, integrating a modular cryostat with custom-designed LEKID arrays.

This paper presented the design and functionalities of the Prime-Cam readout software. 
The software architecture leverages a distributed approach with intelligent RFSoC boards for real-time data acquisition. 
Prioritization of speed is achieved through UDP for time stream data and offloading packet generation to the FPGA fabric.  
A publish-subscribe messaging system ensures efficient communication within the Control Network, while the Timestream Network, designed for high-bandwidth, utilizes UDP for raw detector data transfer. 
This separation of concerns guarantees smooth operation for both control and data acquisition.

To optimize signal-to-noise ratio, the readout system meticulously generates digital waveforms tailored to the specific detector configuration, electrical characteristics, and environmental conditions. 
An iterative process involving broad and targeted sweeps refines the characterization of resonant frequencies, ultimately leading to the creation of an optimal waveform for driving the detector arrays.

Prime-Cam will employ the $IQ$-angle method for calibration, addressing the non-linearity of the detector response. 
This novel implementation utilizes target sweep data and observed detector response to determine the loading-induced frequency deviation, providing an accurate conversion function for extended observation periods, while maintaining required computational efficiency \cite{Gao2008}. 
Additionally, the readout system accounts for temperature sensitivity and optimizes electronics gain settings to maximize the signal-to-noise ratio.

In conclusion, the Prime-Cam readout software offers a robust and scalable solution for complex astronomical instrumentation using KIDs. The system prioritizes efficient communication, real-time data acquisition, meticulous waveform generation, and robust calibration techniques, paving the way for sensitive and groundbreaking submillimeter astronomy surveys with FYST.


\subsection* {Code} 

To facilitate reproducibility and collaboration,
the source code for this research is publicly available on Github: \\
\href{https://github.com/TheJabur/primecam_readout}{https://github.com/TheJabur/primecam\_readout} \\
This allows interested readers to inspect and adapt the code for their own purposes, furthering the scientific dialogue and potential applications of this work.

\subsection* {Acknowledgments}
The CCAT-prime project, FYST and Prime-Cam instrument have been supported by generous contributions from the Fred M. Young, Jr. Charitable Trust, Cornell University, and the Canada Foundation for Innovation and the Provinces of Ontario, Alberta, and British Columbia. The construction of the FYST telescope was supported by the Gro{\ss}ger{\"a}te-Programm of the German Science Foundation (Deutsche Forschungsgemeinschaft, DFG) under grant INST 216/733-1 FUGG, as well as funding from Universit{\"a}t zu K{\"o}ln, Universit{\"a}t Bonn and the Max Planck Institut f{\"u}r Astrophysik, Garching.
Prime-Cam:
The construction of EoR-Spec is supported by NSF grant AST-2009767. The construction of the 350 GHz instrument module for Prime-Cam is supported by NSF grant AST-2117631.
Chai:
The CHAI instrument is supported by DFG grant CRC 956/3, project ID 184018867 as well as funding from Universit{\"a}t zu K{\"o}ln.


\bibliography{PrimeCam_Readout} 
\bibliographystyle{spiejour}

\appendix

\section{Functional Software Overview}

\subsection{Control Computer / \texttt{queen.py}}
\label{sec:queen}

The primary job of the control computer is to facilitate requests through an interface (from the PCS, or through the CLI or local GUI), send commands to the boards, and receive relevant status and diagnostic data back from them.
The main control computer script is \texttt{queen.py}, and this is utilized by the CLI interface script, \texttt{queen\_cli.py}, the GUI interface script, \texttt{queen\_gui.py}, or the PCS agent script, \texttt{queen\_agent.py}.
The control computer also runs a number of daemons to monitor the systems and process return data from the boards.
For example, the queen script is run in `listen' mode as a daemon in order to intercept and data sent back from the boards, primarily for diagnostics.
The control computer also monitors the number of active drones connected and will actively attempt to restart any drones which have gone dark (as well as alert control staff).

The control system is highly-parallel in that a single control computer can control practically any number of realistically deployable boards, and any computer running \texttt{primecam\_readout}, including RFSoC boards, can operate as the control computer. 
This is achievable because the communication medium, \texttt{Redis}, allows for thousands of concurrent connections and is extremely data-efficient when communicating. 
The upper limit to the number of boards a single control computer can control is untested but the theoretical limit is contingent upon factors such as network latency, \texttt{Redis} server capacity, and the inherent communication overhead associated with managing a growing number of parallel boards. 
As the number of parallel boards increases, the potential for contention over shared resources and increased communication traffic may introduce bottlenecks, leading to diminishing returns on system performance. 
Additionally, the computational and memory demands on the centralized control computer can impose practical constraints on the scalability of the system. 
However, the default maximum allowable connections for \texttt{Redis} is 10,000 boards (or approximately 40,000,000 KIDs), and since a single command is extremely bandwidth-light, this is a suggested order for the maximum scalability per control computer.

The control computer connects to both the PCS network and the control network. 
The PCS network allows control through the PCS interface, an API which exposes high-level functionality. 
This API is potentially attached from any observatory control software and allows for functionality such as triggering calibration and data streaming from the detectors. 
These high-level functions are wrappers around lower-level commands to create required workflows.

Low-level commands are exposed in the command-line interface (CLI) and graphical-user interface (GUI). These interfaces are only accessible to local users (or remote users attached via SSH or similar), and are primarily used for development, testing, installation, and diagnostics.
A list of low-level commands can be seen in Table~\ref{tab:commands}.

The CLI accepts command-line arguments to send commands to the boards. Each call can include a single command (plus arguments), but it can be sent to any or all boards.
The GUI also allows sending a single command to any or all the boards, and also includes feedback visuals, for example illustrating the number of current drones operating, and displaying a plot of the last sweep sent from a drone.

\texttt{Queen.py} also contains functionality not directly exposed by the current interfaces, but usable by them. This includes functionality to retrieve a list of commands and convert between command numbers and names, plus low-level functions used by exposed commands.

A range of command numbers is reserved for testing functions. The current software includes a number of automated testing functions which were used in the development and characterization of the instrument, and which serve as examples for further testing function development. These functions can be exposed as commands to be used by an interface, or used directly by a purpose-built interface.

\subsection{Redis}
\label{sec:redis}

\texttt{Redis}, short for Remote Dictionary Server, is an an open-source in-memory  key-value data store that has become an important player in the realm of high-performance data storage and retrieval. 
Initially developed by Salvatore Sanfilippo in 2009, the project was motivated by the need for a high-performance, scalable, and versatile solution for data storage and retrieval in modern applications. \texttt{Redis} was designed to provide a simple yet powerful key-value store that could handle a variety of data structures and offer exceptional speed by keeping the entire data set in memory. \texttt{Redis} has gained widespread adoption due to several key factors that set it apart in the realm of data storage including in-memory storage, data structure versatility, efficient key-value operations, scalability, and high-availability.

\texttt{Redis} follows a client-server architecture, where clients and servers communicate over a network. In this model, the \texttt{Redis} server acts as a central repository for data, and clients, which can be applications or other services, connect to the server to perform various operations. The communication between clients and the server is typically achieved through a protocol such as RESP (REdis Serialization Protocol).
Clients can issue commands to the server, which responds with the requested data or performs the specified operation. \texttt{Redis} supports a wide range of clients written in various programming languages, contributing to its versatility and ease of integration into diverse application stacks.
The client-server model in \texttt{Redis} is designed to be lightweight and efficient, allowing multiple clients to interact with the server concurrently. Clients can subscribe to channels for pub/sub messaging, execute transactions, and leverage various data structures provided by \texttt{Redis}.

We use \texttt{Redis} in a client-server pub/sub model with the control computer running the server, and each of the drones (four per board) running a client. 
Commands are published from the server via custom \texttt{Redis} channels (a unique channel per command) and delivered to subscribed  drones who then act on them.
A drone only intercepts commands intended for it.
The drones are then able to send data and messages back to the control computer.
This is the primary control medium and is extremely light-weight, robust, scalable, and versatile. In addition to the pub/sub communication, \texttt{Redis} also provides status flags and client monitoring, both of which are used to provide additional diagnostic information.

\subsection{RFSoC / \texttt{drone.py}}
\label{sec:drone}

Each board is a Xilinx ZCU111 Radio Frequency System-on-Chip (RFSoC) with expansion ports to enable it to drive 4 RF networks and produce the appropriate timestreams to the data acquisition computer as well as communicate with the control computer. 
The RFSoC has a custom firmware design, and utilizes a custom compiled PYNQ v2.6 image.
Each board is housed in a custom 1U rack-mounted case with liquid cooling and front accessible ports.
Each RF network requires two (frequency multiplexed) coaxial cables into the cryostat, and each board has one control network ethernet cord and one timestream network ethernet cord.
The racks also have individually controllable power supplies, to allow for remote hard reset of any given board if necessary.

Each board runs up to four instances of \texttt{drone.py} where each drone establishes an RF network of up to 1000 KIDs. 
\texttt{drone.py} acts as a mediator between the \texttt{Redis} client and the command functionality script (\texttt{alcove.py}) and manages aspects such as establishing a \texttt{Redis} client connection, logging, managing configuration options, setting and getting key/vals, receiving and parsing commands and their associated arguments, and publishing function return data.

\subsection{Functionality / \texttt{alcove.py} and \texttt{alcove\_commands} Library}
\label{sec:alcove}

The main board functionality is established in \texttt{alcove.py} which imports the \texttt{alcove\_commands} custom library and combines and makes accessible all of the commands which are contained in the library. 
The full list of commands and a short description can be seen in Table~\ref{tab:commands}.

\begin{table}[ht]
\caption{Commands.}
\label{tab:commands}
    \begin{center}
    \begin{tabular}{|c|l|p{65mm}|}
        \hline
        Number & Name & Short description \\
        \hline
        \multicolumn{3}{l}{\textbf{Queen Commands}} \\
        \hline
        1 & alcoveCommand & Directly send command to board[s]. \\
        2 & listenMode & Continuously run and listen for boards return data. \\
        3 & getKeyValue & Get a key-value. Primarily used for status flags. \\
        4 & setKeyValue & Set a key-value.\\
        5 & getClientList & List of every client connected (i.e. drones). \\
        \hline
        10-20 & & Reserved for testing and development functions. \\
        \hline
        \multicolumn{3}{l}{\textbf{Board Commands}} \\
        \hline
        20 & setNCLO & Set the LO frequency. \\
        21 & setFineNCLO & Fine adjustments to the LO. \\
        25 & getSnapData & Capture data from the ADC. \\
        30 & writeTestTone & Send a single tone (at 50 MHz). \\
        31 & writeNewVnaComb & Write a full bandwidth tone comb. \\
        32 & writeTargCombFromVnaSweep & Send a resonator targeted comb. \\
        33 & writeTargCombFromTargSweep & Send a resonator targeted comb. \\
        34 & writeCombFromCustomList & Send a custom comb (from file). \\
        35 & createCustomCombFilesFromCurrentComb & Create custom comb files. \\
        36 & modifyCustomCombAmps & Modify custom comb amplitudes file values by multiplicative factor. \\
        37 & writeTargCombFromCustomList & Calls 34 and also writes required targetted sweep files. \\
        40 & vnaSweep & VNA (full bandwidth) sweep. \\
        42 & targetSweep & Perform a resonator target sweep. \\
        50 & findVnaResonators & Analyse VNA sweep for resonators. \\
        51 & findTargResonators & Analyse target sweep for resonators. \\
        55 & findCalTones & Analyse targetted sweep to find good calibration tone placement. \\
        \hline 
    \end{tabular}
    \end{center}
\end{table}

\subsection{Usage}
\label{sec:usage}

The primary method for installation of the \texttt{primecam\_readout} software during development was through a git clone of the master repository. For Prime-Cam, the PCS and other software operated out of Docker containers, so towards the end of development \texttt{primecam\_readout} was also moved into a Docker container and integrated into the entire software stack. The same software base is installed on both the control computer and the boards. After a git clone operation, the only change needed to become operational is to copy the configuration backup templates and modify to fit the environment, primarily setting IP and MAC addresses.

On the control computer, once a network accessible \texttt{Redis} server is running, \texttt{primecam\_readout} can be used through the various interfaces without a constantly running instance. However if return data are desired to be saved on the control computer, for diagnostics or testing for example, then an instance needs to be running in listen mode to collect the returns:
\begin{verbatim}
 @control: python queen_cli.py 2 -q
\end{verbatim}

\noindent Each drone (up to 4) on every board needs a separate running instance:
\begin{verbatim}
 @board: python drone.py [drid]
\end{verbatim}

\noindent Once the desired drones are running, commands can be sent to it individually by appending its board and drone identifier, or en masse by leaving [bid.drid] blank:
\begin{verbatim}
 @control: python queen_cli.py [com_num] [-a args_str] [bid.drid]
\end{verbatim}
\begin{quote}
[com\_num] is the number of the command to be executed, either a queen command or a board command.

[bid.drid] indicates the board and drone identifiers to send the command to. If only [bid] is given then the command is sent to all drones on that board. If neither a [bid] nor a [drid] is given then the command is sent to all boards and all drones. Note that it does not accept only a [drid].

[-a args\_str] are the arguments to send as parameters for the command in a string format similar to `arg1=val1, arg2=val2'. 

[-q] indicates this is a queen command, as opposed to a board command, as a safety precaution.
\end{quote}

A typical usage scenario of sending the commands to all drones for setting the LO to 500-MHz, performing a full band sweep and roughly identifying the resonator locations, performing a targeted sweep on those locations to identify their resonances within the necessary frequency resolution, and finally setting the tone comb to produce time streams on those resonators, would be as follows:
\begin{verbatim}
 @control: python queen_cli.py 20 -a 500
 @control: python queen_cli.py 31
 @control: python queen_cli.py 40
 @control: python queen_cli.py 50
 @control: python queen_cli.py 32
 @control: python queen_cli.py 42
 @control: python queen_cli.py 51
 @control: python queen_cli.py 33
\end{verbatim}

\noindent Common command sequences can easily be combined together for higher-level commands if desired.

\end{spacing}
\end{document}